\documentclass[showpacs,preprintnumbers,amsmath,amssymb]{revtex4}

\ifx\pdfoutput\undefined
\usepackage{graphicx}
\else
\usepackage[pdftex]{graphicx}
\usepackage{amsmath}
\usepackage{amssymb}
\usepackage{epstopdf}
\fi

\usepackage{dcolumn}
\usepackage{bm}

\newcommand{\be}{\begin{equation}}
\newcommand{\ee}{\end{equation}}
\newcommand{\bear}{\begin{eqnarray}}
\newcommand{\eear}{\end{eqnarray}}
\newcommand{\ba}{\begin{array}}
\newcommand{\ea}{\end{array}}

\newskip\humongous \humongous=0pt plus 1000pt minus 1000pt

\newif\ifdtup

\def\oldreffmt#1{\rlap{[#1]} \hbox to 2\parindent{}}

\def\figfmt#1{\rlap{Figure {#1}} \hbox to 1in{}}

\def\beq{\begin{equation}}
\def\eeq{\end{equation}}
\def\bea{\begin{eqnarray}}
\def\eea{\end{eqnarray}}

\def\bq{\begin{quote}}
\def\eq{\end{quote}}

\relax

\newdimen\tdim
\tdim=\unitlength
\def\bar{\overline}

\begin{document}
\setcounter{page}{0}

\begin{flushright}
ANL-HEP-PR-08-78\\ 
NUHEP-TH/08-11\\
\end{flushright}

\title{Manifestations of Top Compositeness at Colliders}

\author{ {Kunal Kumar$^{a,b}$, Tim M.P. Tait$^{a,b}$, and Roberto Vega-Morales$^{a,b}$}\\[0.5cm]
\normalsize{$^{a}$ Northwestern University, 2145 Sheridan Road, Evanston, IL 60208.\\
\normalsize{$^{b}$ HEP Division, Argonne National Laboratory, Argonne, IL 60439.}}}

\begin{abstract}
We explore the possibility that the right-handed top quark is composite, identifying possible signatures of compositeness and how they might manifest themselves at the LHC and Tevatron.  We perform a complete analysis of the dimension six modifications of the top coupling to gluons and find that cancellations among operators in the $t \bar{t}$ rate allow for very low compositeness scales, but this can be drastically
improved by looking at kinematic distributions.  Turning to the LHC, we examine four top production from
a dimension six four-top operator and estimate the LHC with 100 ${\rm fb}^{-1}$ collected luminosity to be sensitive to compositeness scales as high as 5 TeV.
\end{abstract}

\maketitle
\thispagestyle{empty} 

\section{Introduction}
\label{sec:intro}

With the advent of the Large Hadron Collider era, it is important to understand in detail how 
physics beyond the Standard Model might manifest itself in LHC data.  While detailed exploration 
of the impact of models such as supersymmetric ones is well underway, it is also important to be 
ready for surprises.  The LHC energies are sufficiently beyond any we have previously explored 
directly that there is room for new phenomena that may not be directly connected to outstanding 
theoretical problems such as the stability of the electroweak scale.  In this article, we explore the idea
that the top quark is composite, composed of some new constituent particles 
(``preons") bound together by a new
confining force \cite{Georgi:1994ha,Lillie:2007hd,Pomarol:2008bh}.  As the most recently discovered ingredient 
of the Standard Model, the top is the quark about which we
currently have the least information.  This opens the door to the possibility of large new effects in its 
interactions being relatively weakly constrained by existing measurements, and its large mass may 
be a clue that it is different from other Standard Model fermions.  In fact, many
models exist in which the top quark is composite \cite{Strassler:1995ia}
in order to motivate its large mass, and lead to a variety of interesting signals at the 
LHC \cite{Agashe:2006hk}.

We choose to work generically in a framework in which the right-handed top is composite, without
getting attached to any specific model.  Our hope is to identify interesting phenomena and features
which are not specifically linked to any particular model, but might reasonably be expected to occur
in a broad class of models in which the top is composite.   We use effective field theory as
a framework with which we can explore the most important operators which can modify the top's interactions
with itself and with other particles of the Standard Model.  As a result, the bounds we find can in principle
be applied to any theory which modifies the top's interactions, and are not limited to a specific theory of
top compositeness, or indeed a theory of top compositeness at all.

This article is organized as follows.  In Section~\ref{sec:operators}, we present the most important dimension
six operators leading to modifications of top quark interactions.  They are expected to lead to the dominant
effects at ``low" energies, below the scale of the new physics.  In Section~\ref{sec:ttbar}, we revisit their
effects on $t \bar{t}$ production at the Tevatron.  This extends previous results \cite{Lillie:2007hd}, 
which strictly assumed coefficients as dictated by naive dimensional analysis (NDA)
\cite{Manohar:1983md} to a general effective theory framework, which flexibly describes the effects of heavy new
physics on $t \bar{t}$ production.  As a result, it can describe theories of top compositeness which deviate from
NDA expectations, or indeed {\em any} theory which predicts a modification of the top's coupling to gluons.  We also examine kinematic distributions and find that they are much more powerful than the rate itself in revealing the presence of new physics.  In Section~\ref{sec:LHCtops} we examine the
possibility that the operators modifying the properties of top lie at some what higher energy scales, and deduce their contribution to the production rate of four top quarks at the LHC.  We conclude in 
Section~\ref{sec:conclusions}.

\section{Top Operators Describing New Physics}
\label{sec:operators}

We parameterize the modifications of top interactions through heavy new physics in the form of higher
dimensional operators which we add to the Standard Model action  \cite{Eichten:1983hw}.  
This is justified based on the fact that top observables to date do not show any radical
deviation from the SM predictions and it is reasonable to expect that the scale of compositeness is 
probably somewhat higher than the typical energies probed.   Because at the current time and
in the near future our information about top comes exclusively from hadron colliders, we are most interested in
operators capturing modifications to the interactions of the top quark with itself and/or light quarks 
and gluons, since these lead to the largest effects at a hadron machine.   
The relevant terms in the effective Lagrangian are,
\bea
{\mathcal L} & = & {\mathcal L}_{SM} + \frac{1}{\Lambda^2} \left\{
{\mathcal O}_t + {\mathcal O}_1 + {\mathcal O}_2 + {\mathcal O}_3+ ...  \right\} + {\mathcal O} \left( \frac{1}{\Lambda^4} \right)
\eea
where we define each operator ${\cal O}_i$ below, and additional operators with mass
dimension $> 6$ will have suppressed 
contributions to the processes we consider, which take place at energies $\lesssim \Lambda$.

If $t_R$ and no other Standard Model component is composite, NDA argues that the most significant
operator will involve four top quarks \cite{Georgi:1994ha},
\bea
{\mathcal O}_t & = &
g^2 \left[ \bar{t}^i \gamma^\mu P_R t_j \right] 
\left[ \bar{t}^k \gamma_\mu P_R t_l \right]
\label{eq:4top}
\eea
where $\gamma^\mu$ are the Dirac gamma matrices, $P_R$ is the right-chiral
projector, and $g^2 / \Lambda^2$ is the coupling of this new interaction.  In the compositeness picture,
it can be understood that $g / \Lambda$ represents the amplitude to create the
composite field, and $\Lambda$ itself characterizes the energy scale at which
further elements of the composite sector become important.  The effective theory
is sensible at energies below $\Lambda$ provided $g \lesssim 4 \pi$.
We consider two options for gauge-invariant contractions of the color indices:
contractions of two octets $( T^a )^{j}_i ( T^a )^{l}_k$ and contractions of
two singlets $\delta^j_i \delta^l_k$.
Note that operators constructed by replacing one or more of the top quarks with a
right-handed up or charm quark are also possible, and could lead to more stringent bounds from 
flavor-violating processes.  For simplicity, we assume such operators have small enough coefficients that
they are not important.

While NDA argues that the four top operator is likely to be the most important one in a theory in which the
top is composite, it is very difficult to constrain directly at the Tevatron, where the
energy requirement to make four real top quarks renders the rate negligibly small.  Thus, we turn
to sub-leading operators which can contribute to observable processes, such as
$t \bar{t}$ production.  These operators represent the first potential glimpse that the top is not point-like, as
probed by short wavelength fundamental particles such as gluons.  The three dimension six operators are \cite{Buchmuller:1985jz,Hill:1993hs},

\bea
\mathcal{O}_{1} & = & g_1 g_S
\left[ \left( H \bar{Q}_3 \right) \sigma^{\mu \nu} \lambda^a P_R t \right] G^a_{\mu \nu} + H.c.
\nonumber \\[0.1cm]
\mathcal{O}_{2} & = & g_2 g_S
\left[ \bar{t} \gamma^\mu \lambda^a D^\nu P_R t \right] G^a_{\mu \nu} + H.c.
\nonumber \\[0.1cm]
\mathcal{O}_{3} & = & g_3 g_S
\left[ \bar{t} \gamma^\mu \lambda^a P_R t \right] ~ \sum_q
\left[ \bar{q} \gamma_{\mu} \lambda^a q \right]
\label{eq:ttbaroperators}
\eea
where $g_1$, $g_2$, and $g_3$ parameterize the relative sizes of the three operators, whose over-all
energy scale is set by $\Lambda$, and
we have assumed universal coupling between all light quarks
and top in $\mathcal O_3$.
NDA provides the estimates $g_1(\Lambda) \sim O(1/g)$ and 
$g_2(\Lambda), g_3(\Lambda) \sim O(1)$.  
The operator $\mathcal{O}_{1}$ is a chromo-magnetic moment for 
the top \cite{Atwood:1994vm}, once the Higgs $H$ is replaced with its VEV.
$\mathcal{O}_{2}$ is induced by the four top operator of Eq~(\ref{eq:4top}) through renormalization,
and it was in that limit (with $g_2 (\Lambda)=0$) that the previous study used its influence on
top pair production at the Tevatron to place a limit on top compositeness \cite{Lillie:2007hd}.
The equations of motion relate $\mathcal{O}_{3}$ to $\mathcal{O}_{2} + \mathcal{O}_{2}^\dagger$.  We can thus
consider complex $g_1$ and $g_2$ and $g_3 = 0$ without loss of generality, since a non-zero (real) $g_3$ can
be recast through the equations of motion into a shift in the real part of $g_2$.
For the purposes of our study, we do not restrict ourselves to the values motivated by NDA, allowing for general complex values of $g_1$ and $g_2$.

\section{Top Pairs at the Tevatron}
\label{sec:ttbar}

\begin{figure}
\begin{minipage}[b]{0.5\linewidth}
\centering
\includegraphics[scale=0.3]{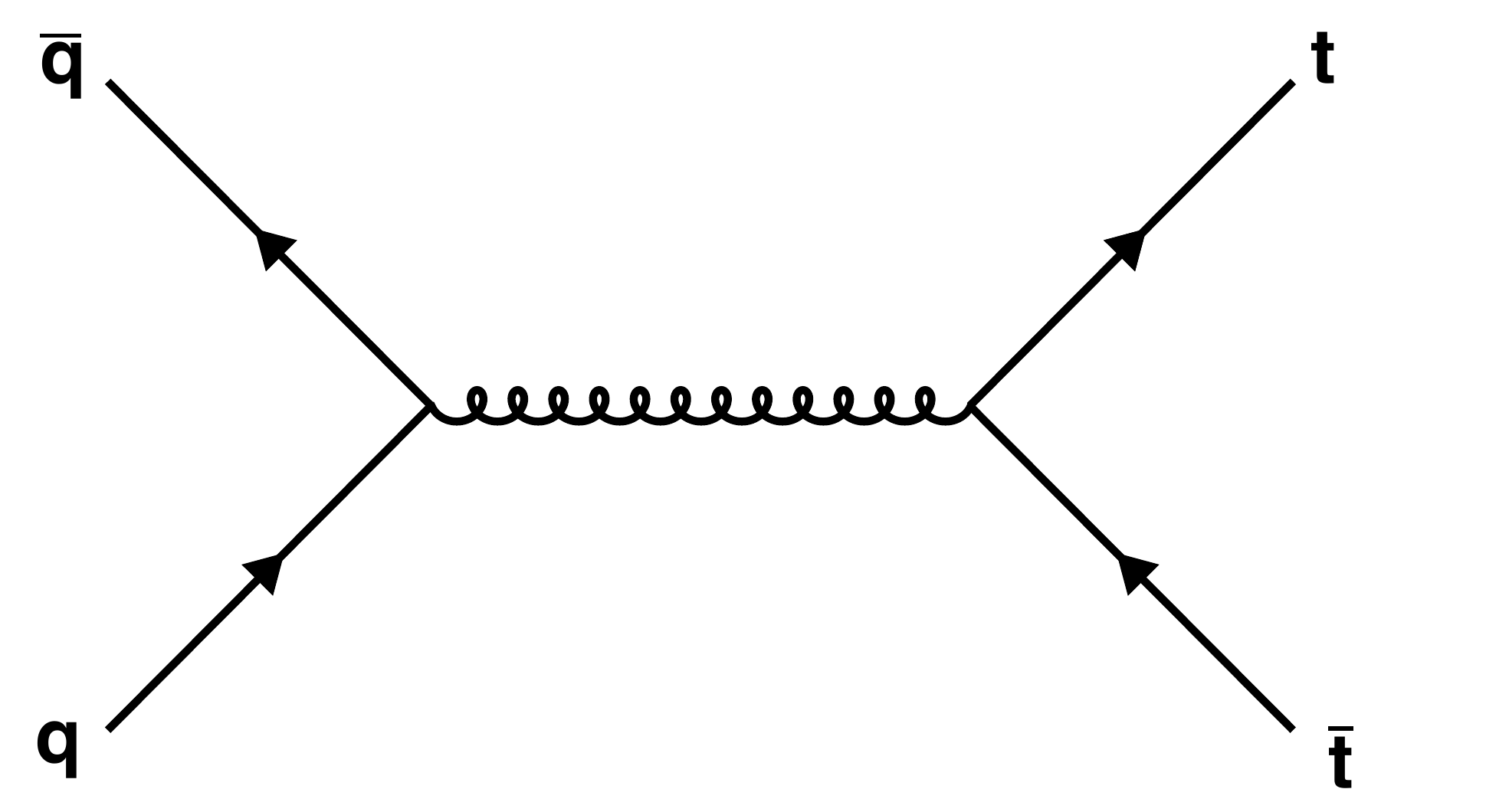} 
\end{minipage}
\hspace{-0.5 cm}
\begin{minipage}[b]{0.5\linewidth}
\centering
\includegraphics[scale=0.28]{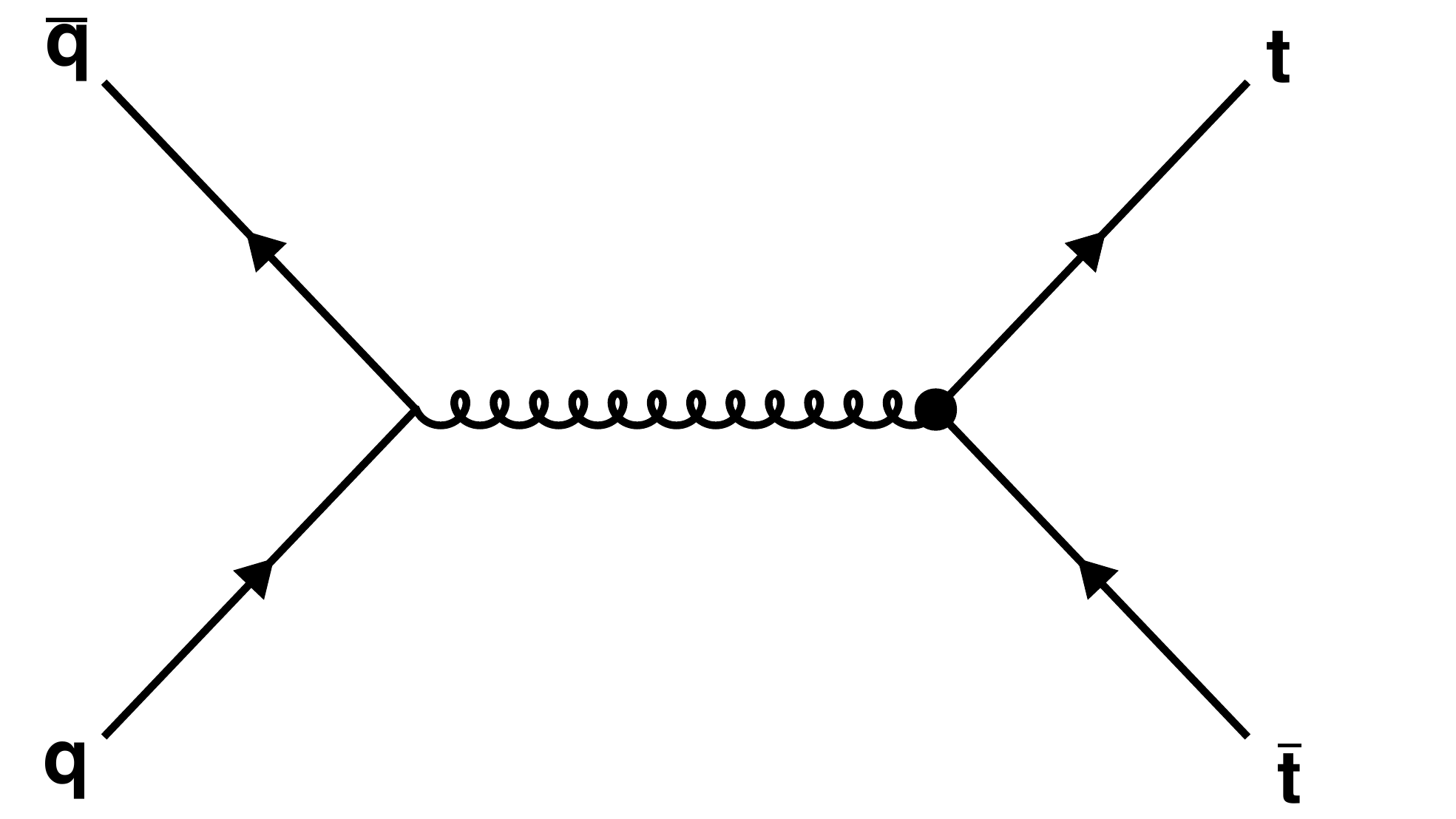} 
\end{minipage}
\caption{\label{fig:Feynman} Feynman diagrams for the process $q\bar{q} \rightarrow t \bar{t}$. 
The shaded blob in the diagram on the right represents the insertion of the new physics 
$g$-$t$-$\bar{t}$ vertex.}
\end{figure}

As alluded above, at the Tevatron the most promising measurements to constrain new physics in the
top system come from $t \bar{t}$ production.  In this section we derive constraints from the inclusive
rate of top pair production, $\sigma (t \bar{t})$.  These constraints turn out to be relatively weak because of 
the possibility of cancellations between the effects of
${\mathcal O}_1$ and ${\mathcal O}_2$, but we show how 
kinematic distributions can improve this situation, and lead to much stronger bounds.  

The inclusive
$t \bar{t}$ cross section is measured by CDF  \cite{CDF-tt}
and D0 \cite{Abazov:2008gc} to be,
\bea
\sigma(t \bar{t})_{CDF} = 7.0 \pm 0.3 \pm 0.4 \pm 0.4~{\rm pb} & ~~~~~~~~ & 
\sigma (t \bar{t})_{D0} = 7.62 \pm0.85 ~\rm{pb} ~,
\eea
(quoted at $m_t = 172.6$ GeV) where the errors are (in order)
statistical, systematic, and arising from the luminosity measurement.
Both are slightly higher than the Standard Model prediction,
\bea
\sigma (t \bar{t})_{SM} & = & 6.6 \pm 0.8~{\rm pb}~,
\eea
(we combine results from both references of \cite{Kidonakis:2003vs},
to obtain this estimate), but not significantly so.  The CDF measurement has slightly smaller error bars,
and is slightly closer to the SM,  and thus results in the stricter bound.
In order to be conservative, we base our limit on it, combining
the various errors in quadrature to arrive at
$\sigma (t \bar{t})_{exp} = 7.0 \pm 0.61$~pb.

At the Tevatron, $q \bar{q}$ initial states provide more than $80\%$ of the rate, and thus the dominant effect
of the operators modifying the $g$-$t$-$\bar{t}$ interaction results from an insertion of the modified vertex in
the usual SM production diagram, as shown in Figure~\ref{fig:Feynman}.  Since SM production does a reasonably good job describing the data, we work at leading order in the $1/\Lambda^2$, in which the modified vertex
interferes with the usual SM production diagram.  We neglect the gluon-initiated graph, which at the Tevatron
amounts to an error of roughly $10\%$ or so in our estimates.  In the composite top theory, this correction looks
like a hard gluon interacting directly with some of the colored preons inside the top, probing its internal structure.
Up to ${O}(1 / \Lambda^{2})$, the partonic cross section $\hat{\sigma} ( q \bar{q} \rightarrow t \bar{t})$ is given by,
\bea
\hat{\sigma} \propto \left| \mathcal{M} \right|^2 = \left| \mathcal{M}_{SM} + \mathcal{M}_{NP} \right|^2 
= \left| \mathcal{M}_{SM} \right|^2 
+ 2 {\rm Re} \left[\mathcal{M}^{*}_{SM}\mathcal{M}_{NP} \right] + O \left( \frac{1}{\Lambda^4} \right) ~,
\eea
with the leading new physics correction arising from the cross term between the standard model and new physics amplitudes. The resulting cross section can be rewritten to be proportional to $\sigma_{SM}$,
\bea
\hat{\sigma} = \hat{\sigma_{SM}} 
\left(1 + ~{\rm Re}~ \frac{ g_1(16vms^2)+g_2(4m^2s^2+s^3+s(s+2t-2m^2)^2}{2\Lambda^2(2m^4+s^2-4m^2t+2st+2t^2)} \right) + O \left( \frac{1}{\Lambda^4} \right) ~.
\label{eq:factor}
\eea
where $s$ and $t$ are the Mandelstam variables, and $m$ is the top quark mass.  Note that at leading order,
only the real parts of $g_1$ and $g_2$ contribute.

\begin{figure}
\includegraphics[angle=0,scale=.75]{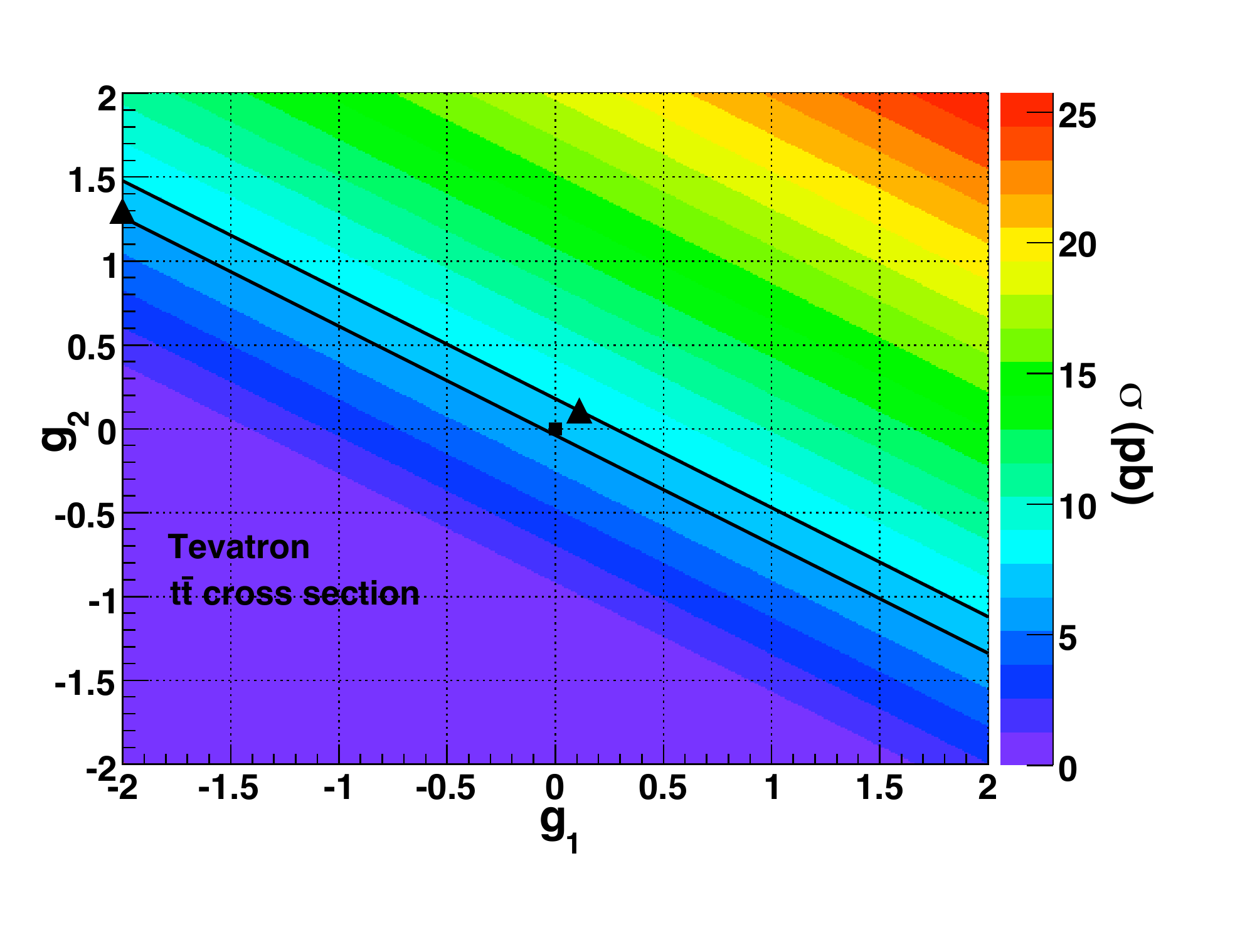} 
\caption{\label{fig:g1g2space}
The inclusive top pair production cross section in the plane of $g_1$ and $g_2$. The region between the 
two black lines is consistent with Tevatron $t\bar{t}$ cross section measurements at $1\sigma$, and 
the Standard Model prediction is represented by the black square. The black triangles denote two points 
which lead to a cross section within $1\sigma$ of the experimental results, and for which we present 
detailed distributions in Figure~\protect{\ref{fig:PT2Mass}}.}
\end{figure}

We simulate the new physics effect by generating SM $t \bar{t}$ events using Madevent \cite{Alwall:2007st}
and subsequently reweighting them by Eq.~(\ref{eq:factor}) on an event-by-event basis, according to the
generated kinematics.  We vary the real parts of $g_1$ and $g_2$, with $\Lambda$ fixed to 500 GeV, and
in Figure~\ref{fig:g1g2space} we present the result for the inclusive $t \bar{t}$ cross section in the
plane of $g_1$ and $g_2$.  The resulting cross section is somewhat more sensitive to ${\mathcal O}_2$ 
than to ${\mathcal O}_1$.  Also shown are the one sigma contours around the central experimental 
measurement, revealing that if $g_1$ and $g_2$ have opposite signs, their effects in $\sigma (t \bar{t} )$
may cancel, resulting in no large net effect in the inclusive rate, despite the presence of low scale new physics.

We expect to better constrain $g_1$ and $g_2$ by comparing the distributions of 
the $t \bar{t}$ invariant mass \cite{Frederix:2007gi}
and top rapidity with experimental data. To illustrate the effectiveness of these distributions in
revealing the presence of physics beyond the Standard Model, we consider two points in the parameter 
space for which we compute the $t \bar{t}$ invariant mass and top rapidity distributions.  We choose our 
two points to lie near the boundary of experimental error, thus leading to the largest deviation from the Standard Model not already in contradiction
with the measured $t \bar{t}$ cross section.  These points are indicated by black triangles on 
Figure~\ref{fig:g1g2space},
\bea
{\rm Point~1}~:~\left( g_1 , g_2 \right) = \left( 0.11, 0.11 \right)
& ~~~~~~~~~~~ & {\rm Point~2}~:~\left( g_1 , g_2 \right) = \left( -2.0, 1.3 \right) ~.
\eea
Point 1 represents a case where both operators do not result in any cancellation in the inclusive rate, whereas Point 2 is a case where moderate cancellations result in a $t \bar{t}$ cross section consistent with the Standard
Model.

\begin{figure}
\begin{minipage}[b]{0.5\linewidth}
\centering
\includegraphics[width=8.6 cm]{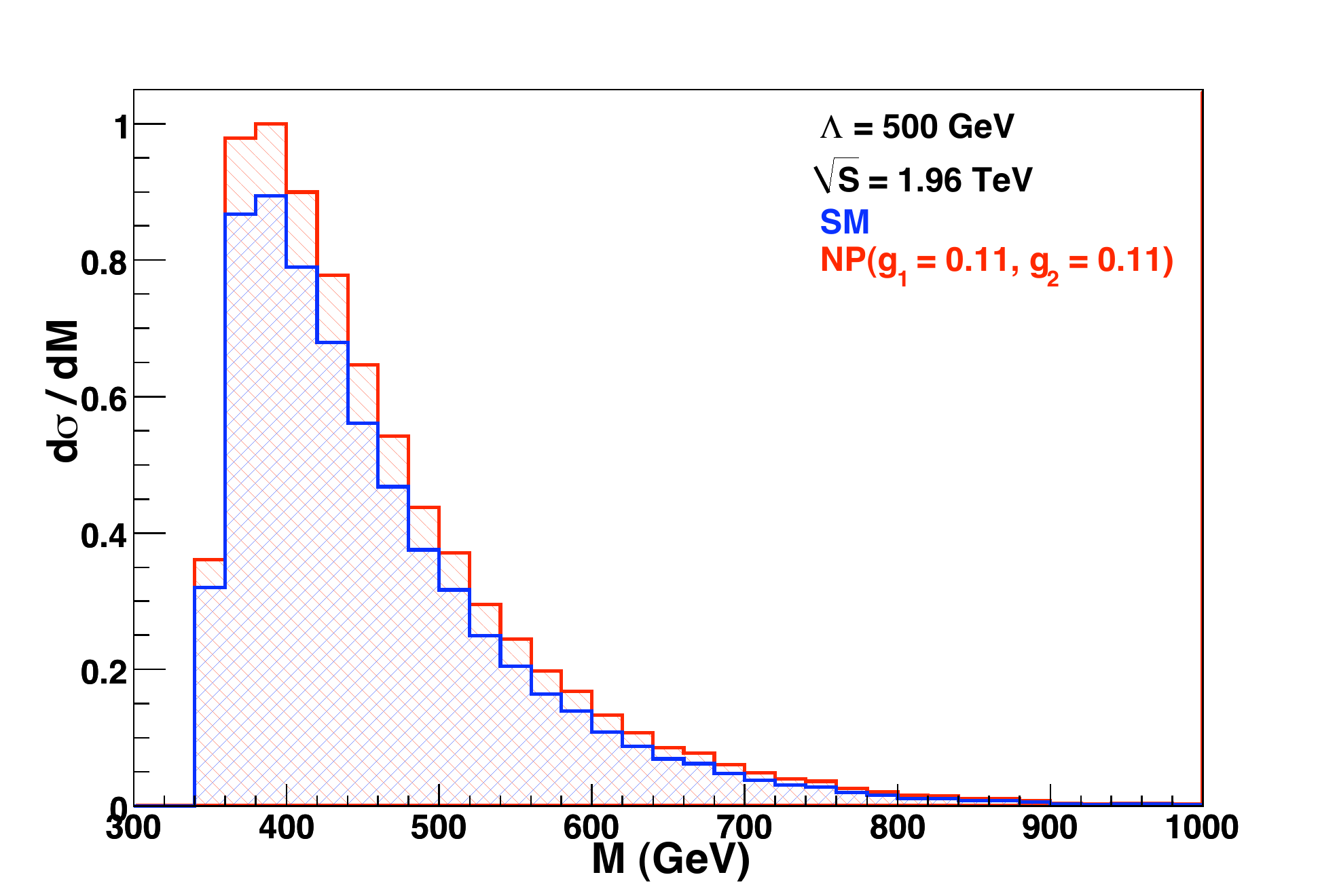} 
\end{minipage}
\hspace{-0.2 cm}
\begin{minipage}[b]{0.5\linewidth}
\centering
\includegraphics[width=8.6 cm]{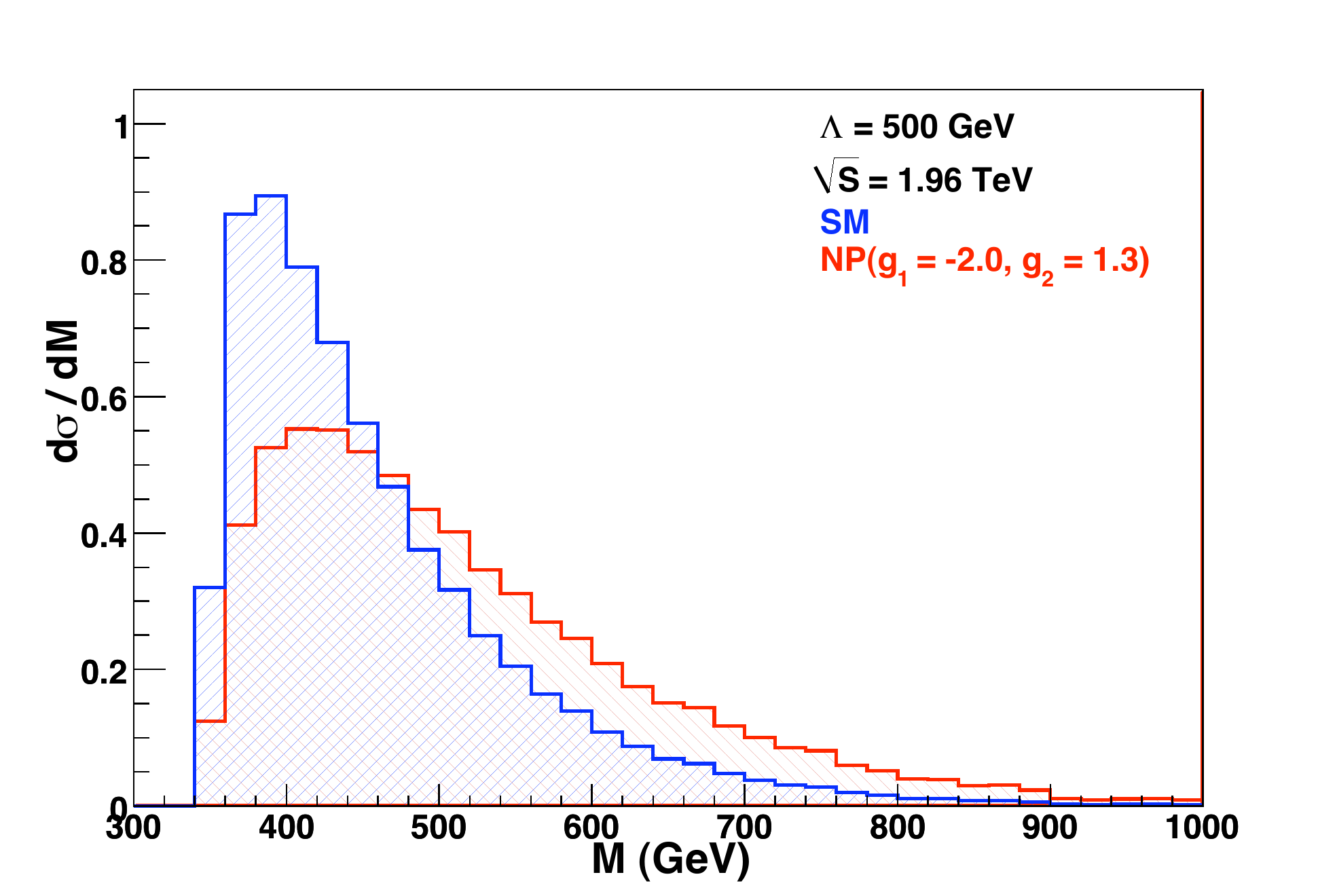} 
\end{minipage}
\caption{\label{fig:PT2Mass} Invariant mass distribution of $t\bar{t}$ for (g1= g2 = 0.11) and (g1 = -2.0, g2 = 1.3) }
\end{figure}

In Figure~\ref{fig:PT2Mass}, we present the invariant mass and rapidity distributions for both points.  Our results
reveal that despite the mild effect on the $t \bar{t}$ rate, the effect of Point 2
on the $t \bar{t}$ invariant mass distribution is striking --
in fact, so striking that this point is certainly already ruled out by the searches for $t \bar{t}$ resonances
\cite{Aaltonen:2007dz}.  While that data is not presented in such a way as to allow an outsider to
perform the analysis, we encourage the experimental collaborations to consider deriving such bounds
as a by-product of the resonance searches.  The effects on the top rapidity distributions are somewhat more
modest, but with the interesting feature that because ${\mathcal O}_2$ distinguishes the top from the anti-top,
it can lead an asymmetric top rapidity distribution, and a deviation in the forward-backward asymmetry of the top
\cite{Aaltonen:2008hc}.

\begin{figure}
\begin{minipage}[b]{0.5\linewidth}
\centering
\includegraphics[width=8.6 cm]{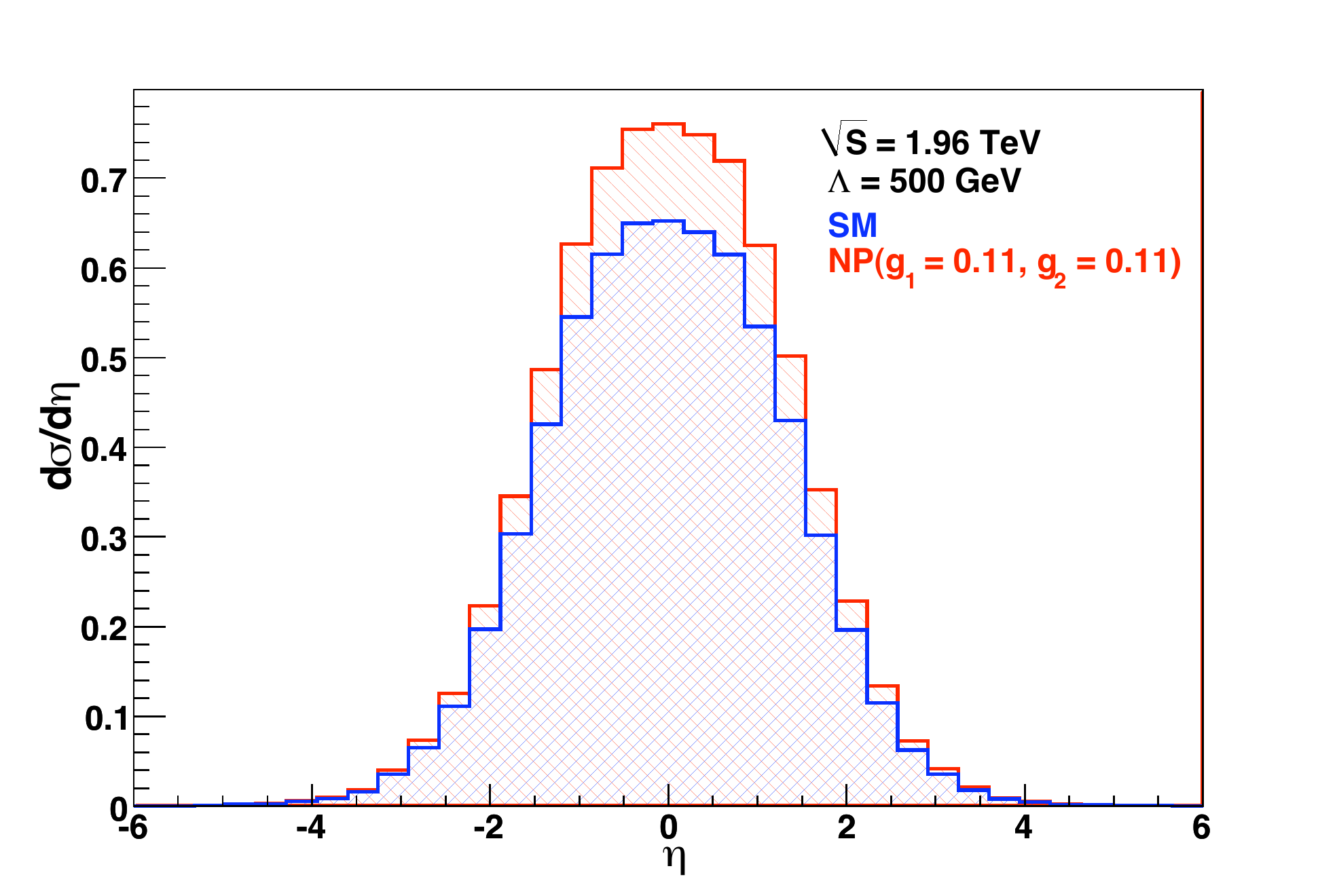} 
\end{minipage}
\hspace{-0.2 cm}
\begin{minipage}[b]{0.5\linewidth}
\centering
\includegraphics[width=8.6 cm]{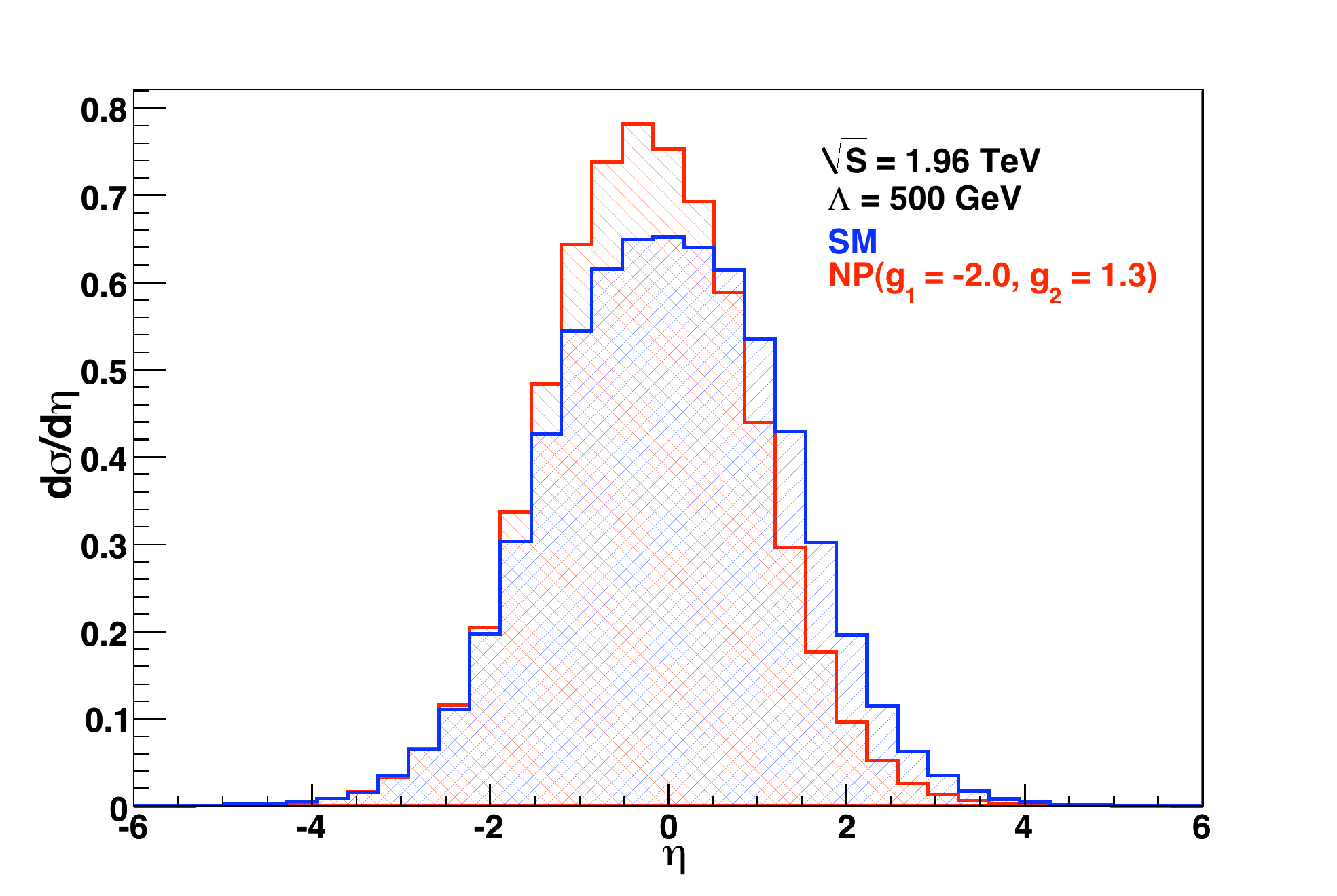} 
\end{minipage}
\caption{\label{fig:PT2Rapidity} Rapidity distribution of the top quark for (g1= g2 = 0.11) and (g1 = -2.0, g2 = 1.3) }
\end{figure}

\section{Four Tops at the LHC}
\label{sec:LHCtops}

\begin{figure}
\includegraphics[angle=0,scale=0.35]{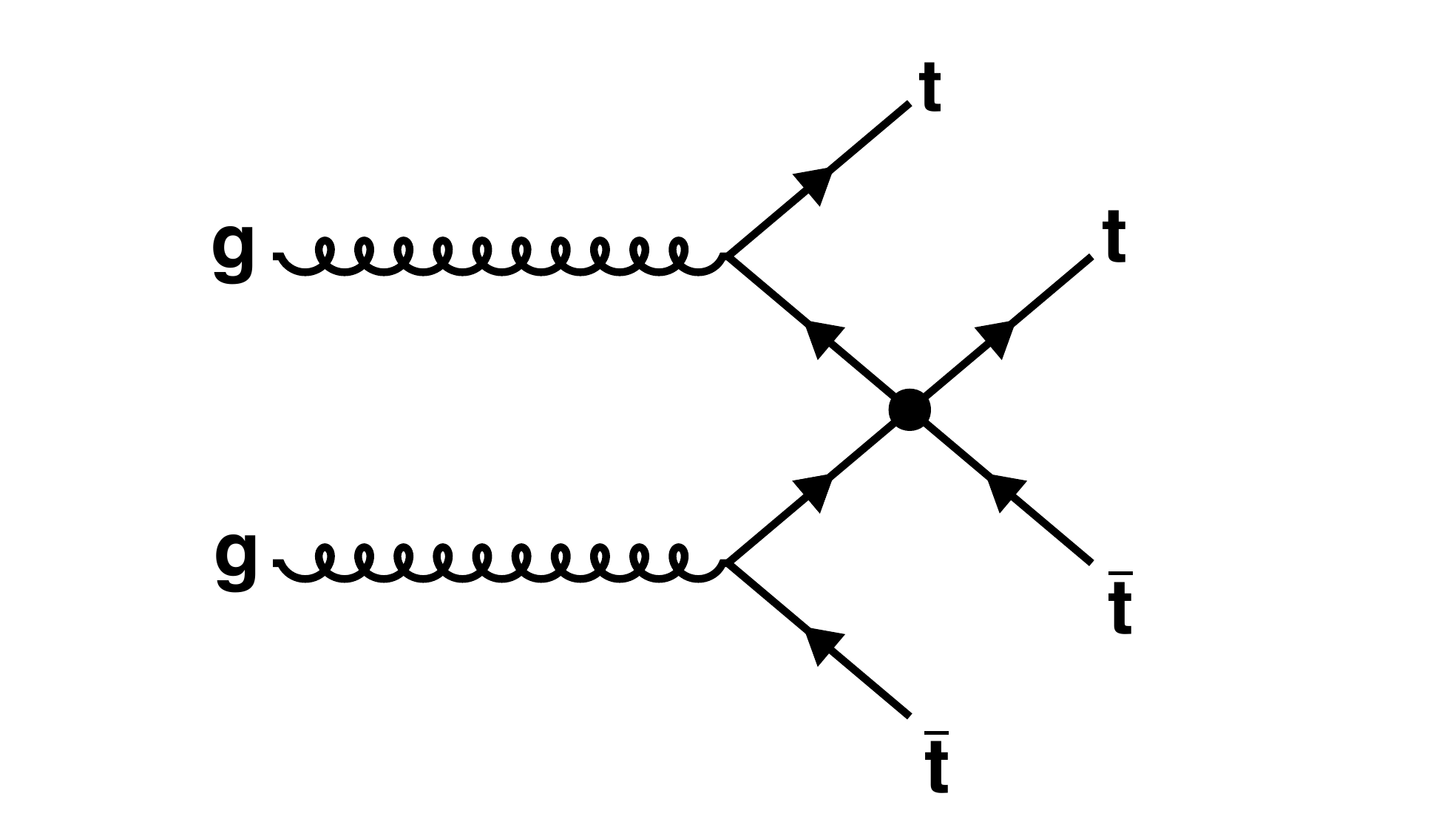} 
\caption{\label{fig:gg4top} Representative Feynman diagram illustrating the contribution from a four top operator 
to four top production at the LHC.}
\end{figure}

At the LHC, energies are sufficiently large to consider processes mediated by the four top operator of
Eq.~(\ref{eq:4top}), namely $pp \rightarrow t\bar{t} t\bar{t}$ through processes such as 
$pp \rightarrow t\bar{t}^*$ followed by $\bar{t}^* \rightarrow \bar{t}t\bar{t}$ (see Figure~\ref{fig:gg4top} for
a representative Feynman diagram).    As four top production is characterized by higher energies 
than $t \bar{t}$, it naturally provides a larger lever-arm with which to study higher dimensional operators.  
In this section, we consider larger values of $\Lambda$, such that the effective theory truncated at order
$1/\Lambda^2$ remains an accurate description at LHC energies.  Thus, for the purposes of this section
we consider the case where the new physics results in effects probably too small to be seen at the Tevatron,
and the LHC is the first place where non-standard interactions for top may be visible. 

Since the Standard Model prediction for four top production at the LHC is very small (perhaps
unobservably so \cite{Lillie:2007hd,Gerbush:2007fe}), any positive signal at the LHC will arise in a regime where the new physics amplitudes for $gg \rightarrow t \bar{t} t \bar{t}$ are much larger than the 
sum of the Standard Model amplitudes.  Thus, for this process to be interesting, we expect that the
$O(1/\Lambda^4)$ contribution to the cross section from the square of diagrams such as the one in
Figure~\ref{fig:gg4top} will dominate over both the Standard Model contribution and
the $O(1/\Lambda^2)$  and $O(1/\Lambda^4)$ interference terms with dimension 8 operators.  We neglect these smaller contributions in presenting our results.
A very similar study (for the octet case) was previously presented in \cite{Pomarol:2008bh}, including interference with the SM amplitudes, and the results are similar to the ones found here.

\begin{figure}
\includegraphics[angle=0,scale=0.7]{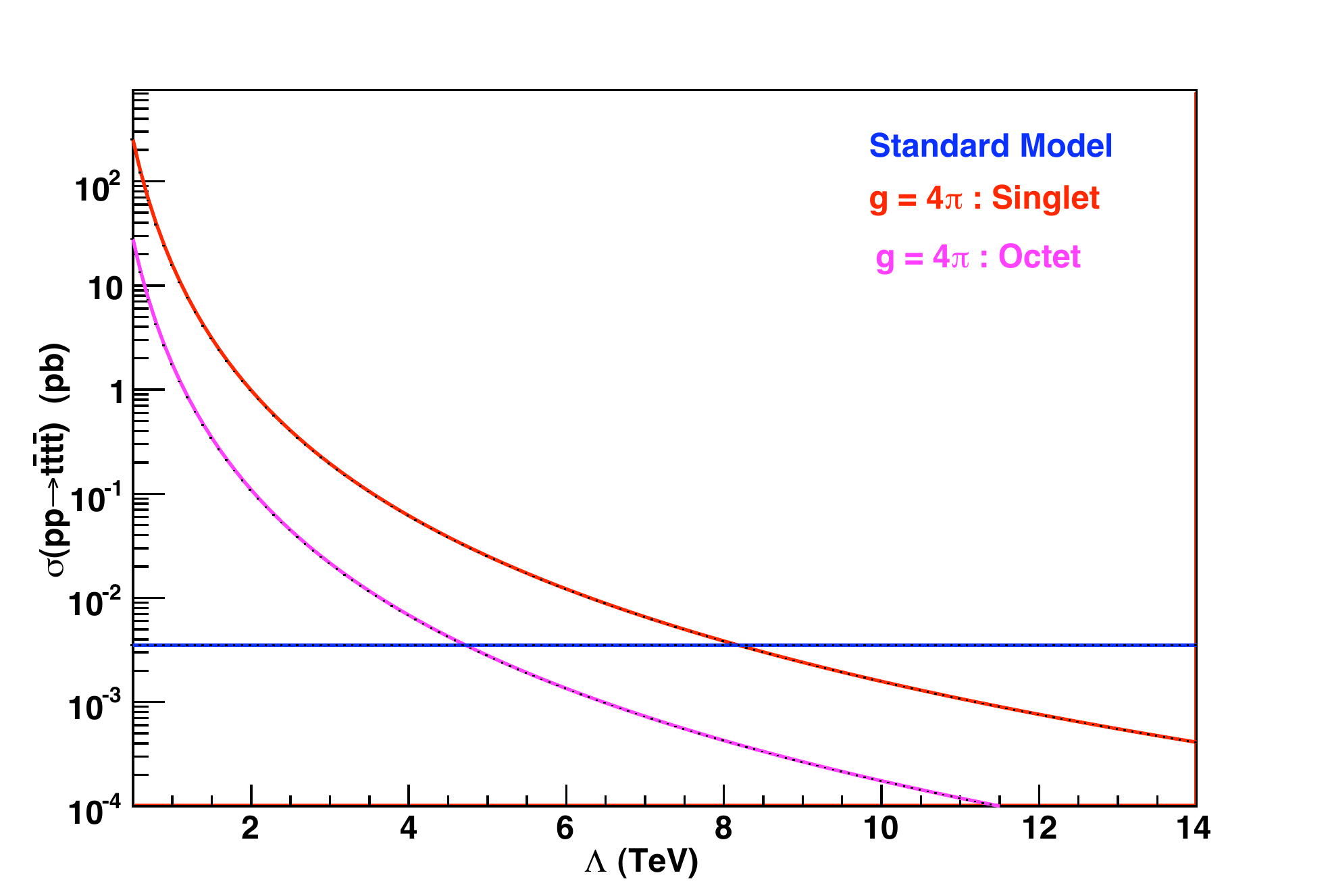} 
\caption{\label{fig:4topcsxn} Rate of four top production at the LHC as a function of $\Lambda$ with
$g=4\pi$.  Also indicated on the graph is the leading order SM prediction for four top production of about
4 fb.}
\end{figure}

We implement the four top operator into MadEvent by introducing an auxiliary field which mimics the
dimension six operator at low energies (we verify that the result scales as expected with the mass of the auxiliary field as a cross check).  For a theory with a composite top, with
strongly coupled new physics, we set $g = 4 \pi$, as suggested by NDA;  the rates
for theories with different values of $g$ are easily obtained by scaling our results by $g^4 / (4 \pi )^4$.
The resulting cross section, as a function of $\Lambda$, is shown in Figure~\ref{fig:4topcsxn}.  

The Standard Model prediction for the four top rate is about 3 fb \cite{Lillie:2007hd},  and we see that new
physics can lead to enhancements by many orders of magnitude over the SM.  A 
previous study \cite{Lillie:2007hd} based on like-sign leptonic $W$ decays and 100~${\rm fb}^{-1}$ of integrated luminosity
concluded that four tops are observable over
backgrounds (dominantly di-boson production plus jets and $t \bar{t}$) provided the raw four top production
is greater than 45 fb.  From Figure~\ref{fig:4topcsxn}, this occurs for $\Lambda \leq 5$~TeV, and is
the largest scale of top compositeness to which the LHC (with 100~${\rm fb}^{-1}$)
has sensitivity based on a simple like-sign lepton analysis.  Since our production is through a 
higher dimensional operator and favors relatively boosted tops, it would be interesting to see
if hadronic top reconstruction through jet shape variables could provide additional channels with which
the four top rate could be measured \cite{Kaplan:2008ie}.

Many other theories can lead to an excess in the four top production rate, including theories
with color octet scalars \cite{Gerbush:2007fe,Dobrescu:2007yp},
color octet vectors \cite{Guchait:2007jd}, color sextets \cite{Chen:2008hh},
and supersymmetric theories with decay chains through third family sfermions \cite{Hisano:2002xq}.
Theories with additional down-type quarks \cite{Choudhury:2001hs}
which decay into $W$ and top could also be revealed by
the like-sign $W$ search strategy.  In each case, there are additional features in the kinematic distributions
which can help distinguish which model is responsible for a positive signal.  For example, we expect to
reconstruct resonances in $t \bar{t}$ pairs for the octets, resonances 
in $tt$ ($\bar{t} \bar{t}$) for the sextets, additional sources of missing energy 
beyond the leptonic $W$ decays in the case of supersymmetric theories, and $t W$ resonances for the
theories with additional fermions.  A systematic study of the characteristics of four top signals would be
interesting, but is beyond the scope of this work.

\section{Outlook}
\label{sec:conclusions}

The top quark is a mystery.  Its large mass may be an indication that is special in some way, and its status as
the most recently discovered ingredient within the Standard Model allows great opportunity for new physics
to cause its properties to deviate from our expectations.  The truth about top is still largely unknown.
In this article, we explore some of the deviations possible
at the Tevatron and LHC when the top is not standard.  

We find that bounds from top pair production at the
Tevatron provide some information, but are relatively weak because of the possibility of cancellations between
contributions from new physics.  Kinematic distributions are much more powerful.  The invariant mass distribution of the $t \bar{t}$ system can be pushed to peak at higher values, and the rapidity distribution may become more
central or shifted asymmetrically, leading to a deviation in the forward-backward charge asymmetry of top pairs.

At the LHC, a more promising channel is provided by the production of four top quarks.  The Standard Model predicts a very small rate for this process, leaving a lot of room for new physics to creep in.  We find that a strongly coupled four top operator can lead to observable deviations even when its characteristic energy scale is multi-TeV.

While we are motivated by the picture that the top quark is composite, our results are presented in the language
of effective field theory, and can be used far beyond the scenario at hand.  {\em Any} theory of new physics modifying the inter-top interactions or top coupling to gluons is described at low energies by these operators.
Once a deviation is found, the pattern of couplings reflects the UV physics which gave rise to the operators and provides the blue-print through which we can build a UV-complete description.

\vspace*{0.5cm}

{\bf Acknowledgments }\\

The authors are pleased to acknowledge conversations with J. Alwall and 
D. Choudhury.
T Tait is grateful to the SLAC theory group for their excellent hospitality and over-all awesomeness
while this work was being completed.
Research at Argonne National Laboratory is 
supported in part by the Department of Energy 
under contract DE-AC02-06CH11357.


\end{document}